\documentclass{mystyle}

\usepackage{graphicx}
\usepackage{amsmath}  

\newcommand{\g}{$\gamma$}
\newcommand{\nhi}{N(\mathrm{H\,\scriptstyle{I}})}

\newcommand{\nhd}{N(\mathrm{H_2})}
\newcommand{\wco}{W_\mathrm{CO}}
\newcommand{\hi}{\mathrm{H\,\scriptstyle{I}}}

\newcommand{\xco}{X_\mathrm{CO}}

\newcommand{\qhi}{q_\mathrm{H\,\scriptscriptstyle{I}}}
\newcommand{\qco}{q_\mathrm{CO}}

\title{The interstellar environment in the outer Galaxy as seen in \g-rays by \textit{Fermi}}
\shorttitle{The \textit{Fermi} view of the interstellar environment in the outer Galaxy}
\author{L.~Tibaldo\from{ins:x}\from{ins:y}\thanks{Partially supported by the International
Doctorate on AstroParticle Physics (IDAPP) program.}\ETC,
I.~A.~Grenier\from{ins:y},
T.~Mizuno\from{ins:z},
on behalf of the \textit{Fermi} LAT collaboration}
\instlist{\inst{ins:x} Istituto Nazionale di Fisica Nucleare, Sezione di Padova, and Dipartimento di
Fisica ``G. Galilei'', Universit\`a di Padova, 35131 Padova, Italy
  \inst{ins:y} Laboratoire AIM, CEA-IRFU/CNRS/Universit\'e Paris Diderot, Service d'Astrophysique,
CEA Saclay, 91191 Gif sur Yvette, France
  \inst{ins:z} Department of Physical Sciences, Hiroshima University, Higashi-Hiroshima, Hiroshima
739-8526, Japan}
\PACSes{
\PACSit{95.85.Pw}{Astronomical observations: \g-rays}
\PACSit{98.38.Am}{ISM in the Milky Way: physical properties}
\PACSit{98.70.Sa}{Cosmic rays}
}
\begin{document}

\maketitle

\begin{abstract}
Gamma-ray emission produced by interactions between cosmic rays (CRs) and interstellar gas traces
the product of their densities throughout the Milky Way. The outer Galaxy is a
privileged target of
investigation to separate interstellar structures seen along the line of sight. Recent observations
by the
\textit{Fermi} Large Area Telescope (LAT) shed light on open questions of the EGRET era about the
distribution of CR densities and
the census of the interstellar medium.

The gradient of \g-ray emissivities measured in the outer Galaxy is significantly flatter than
predictions from widely used CR propagation models given the rapid decline of putative CR sources
beyond the solar circle. Large propagation volumes, with halo heights up to 20~kpc, or a flat CR
source distribution are required to match the data. Other viable possibilities include non-uniform
CR diffusion  properties or more gas than accounted for by the radio/mm-wave data.
\g-ray data constrain the evolution of the $\xco=\nhd/\wco$ ratio within a few kpc from the Sun.
There is a significant increase by a factor~2 from nearby clouds in the Gould Belt to the local
spur. No further significant variations are measured from the local spur to the Perseus spiral arm.
At the level of statistical accuracy provided by the LAT data, the most important source of
uncertainty, often overlooked so far, is due to the optical depth correction applied to derive the column densities
of $\hi$. Reliable determinations of the amount of atomic gas in the plane are key to better probe
the properties of CRs in the Galaxy. 
\end{abstract}

\section{Introduction}

Interstellar \g-ray emission is produced by interactions of high-energy cosmic rays (CRs) with the
gas in the interstellar medium (ISM) and
the soft interstellar radiation fields. Its observations carry information
about CR properties in distant locations and it provides a tracer of the total interstellar gas densities to be compared 
with radio/mm-wave data: the 21-cm line of atomic hydrogen, $\hi$,
and the 2.6-mm line of CO, used as a surrogate tracer of molecular mass.

Open issues in the understanding of the interstellar \g-ray emission concern the identification and
spatial distribution of CR sources and the census of the ISM, notably the $\xco=\nhd/\wco$ conversion
factor. The outer Galaxy is a privileged
observational target since the Doppler shift of radio/mm-wave lines due to the Galactic
rotation unambiguously locates the emitting clouds. There are two longitude windows with a steep
velocity gradient leading to a good kinematic separation \cite{ref:2}.

We reported analyses of recent measurements by the Large Area Telescope (LAT)
on board the \textit{Fermi \g-ray Space Telescope} \cite{ref:4} for the second \cite{ref:3}
and third \cite{ref:5} Galactic quadrants. The component separation based on likelihood fitting
allowed us to extract the emissivities per H atom, $\qhi$, and per $\wco$
unit, $\qco$, in several regions along the lines of sight as described in table~\ref{tab:regions}.
We refer the
interested reader to the aforementioned papers for details about the analysis and we briefly discuss
here the
implications of the results for the distribution of CRs in the Galaxy and the calibration of the
$\xco$ ratio.

\begin{table}[!htbp]
  \caption{Regions seen toward the outer Galaxy and their approximate Galactocentric
distances.}
  \label{tab:regions}
  \begin{tabular}{rcc}
    \hline
			& second quadrant  & third quadrant    \\
			& $100^{\circ}<l<145^{\circ}$ 	& $210^{\circ}<l<250^{\circ}$ \\
    \hline
      Gould Belt	& $8.5-8.8$ kpc	& --		\\
      Local Arm     	& $8.8-10$ kpc	& $8.5-10$ kpc	\\
      interarm region	& --     	& $10-12.5$ kpc	\\
      Perseus arm	& $10-14$ kpc	& $12.5-16$	\\
      outer region	& $>14$ kpc	& $>16$ kpc	\\
    \hline
  \end{tabular}
\end{table}

\section{The spatial distribution of CR densities}

Provided that the $\hi$ column densities are accurately measured from radio data, 
the emissivity $\qhi$, \idest\ the \g-ray emission rate per hydrogen atom, directly relates to the
average CR densities in each of the regions considered.  The $\qhi$ profile (integrated above
200 MeV) as a function of Galactocentric distance $R$ is shown in fig.~\ref{fig:profiles} (left
panel).

CR densities appear to be uniform within 20\% for comparable Galactocentric distances in the two
regions studied. There is a small decrease from nearby complexes in the Gould Belt\footnote{Our
measurement of the local emissivity is compatible with expectations based on CR spectra as they are
measured near the Earth \cite{ref:3,ref:7}.} to the Perseus spiral arm, but no further significant
gradient is observed beyond $R \simeq 11$~kpc.

We assessed the uncertainties due to the optical depth correction applied to
the $\hi$ line intensities to derive $\nhi$. In fig.~\ref{fig:profiles} (left panel) we have explored a range of possible spin temperatures ($T_S$),
showing that the uncertainties in the $\hi$ mass, often overlooked in the past, dominate the
errors in the emissivity determination.

\section{The calibration of molecular masses}
\begin{figure}[!btp]
\begin{center}
\begin{tabular}{cc}
\includegraphics[width=0.5\textwidth]{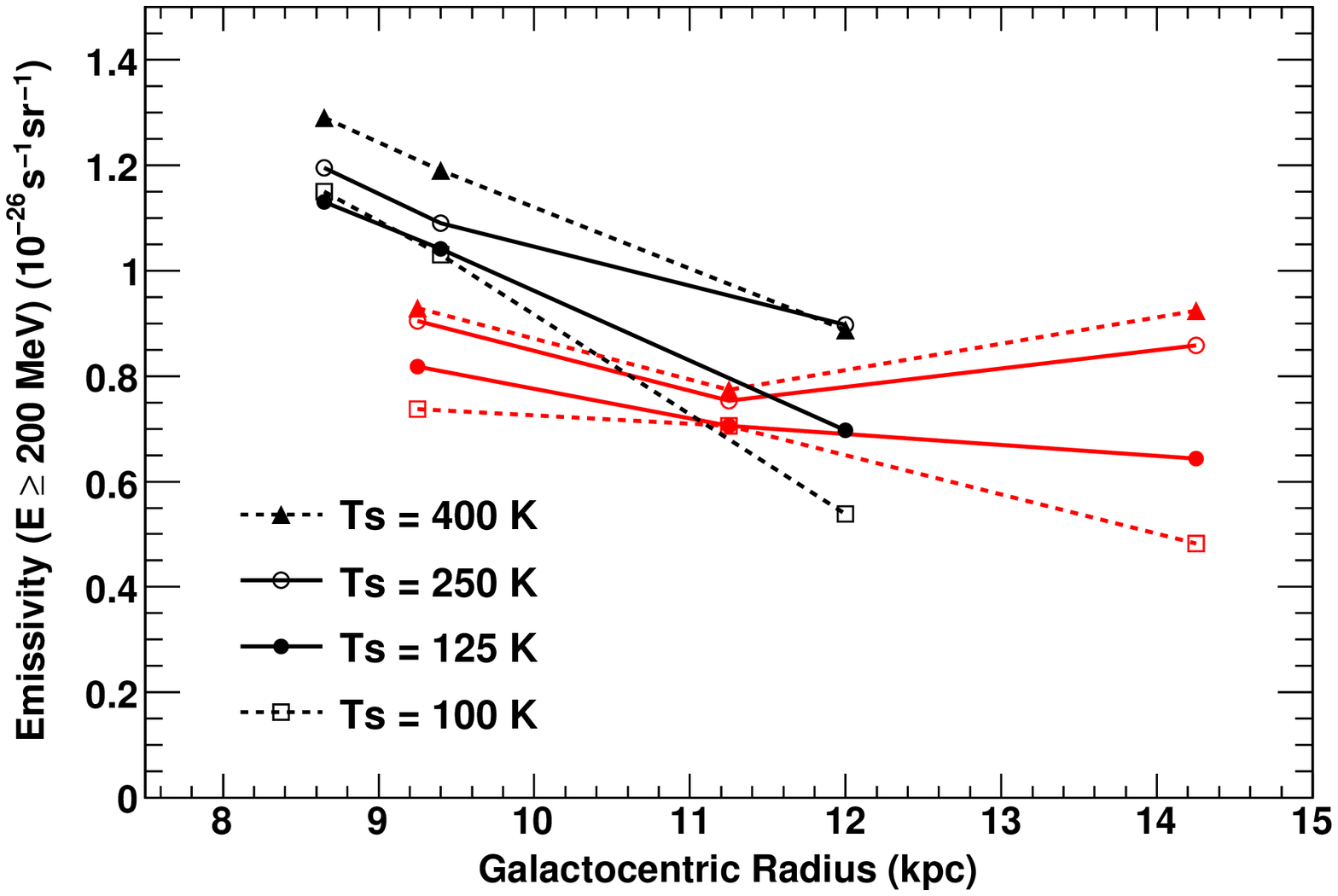}
&
\includegraphics[width=0.5\textwidth]{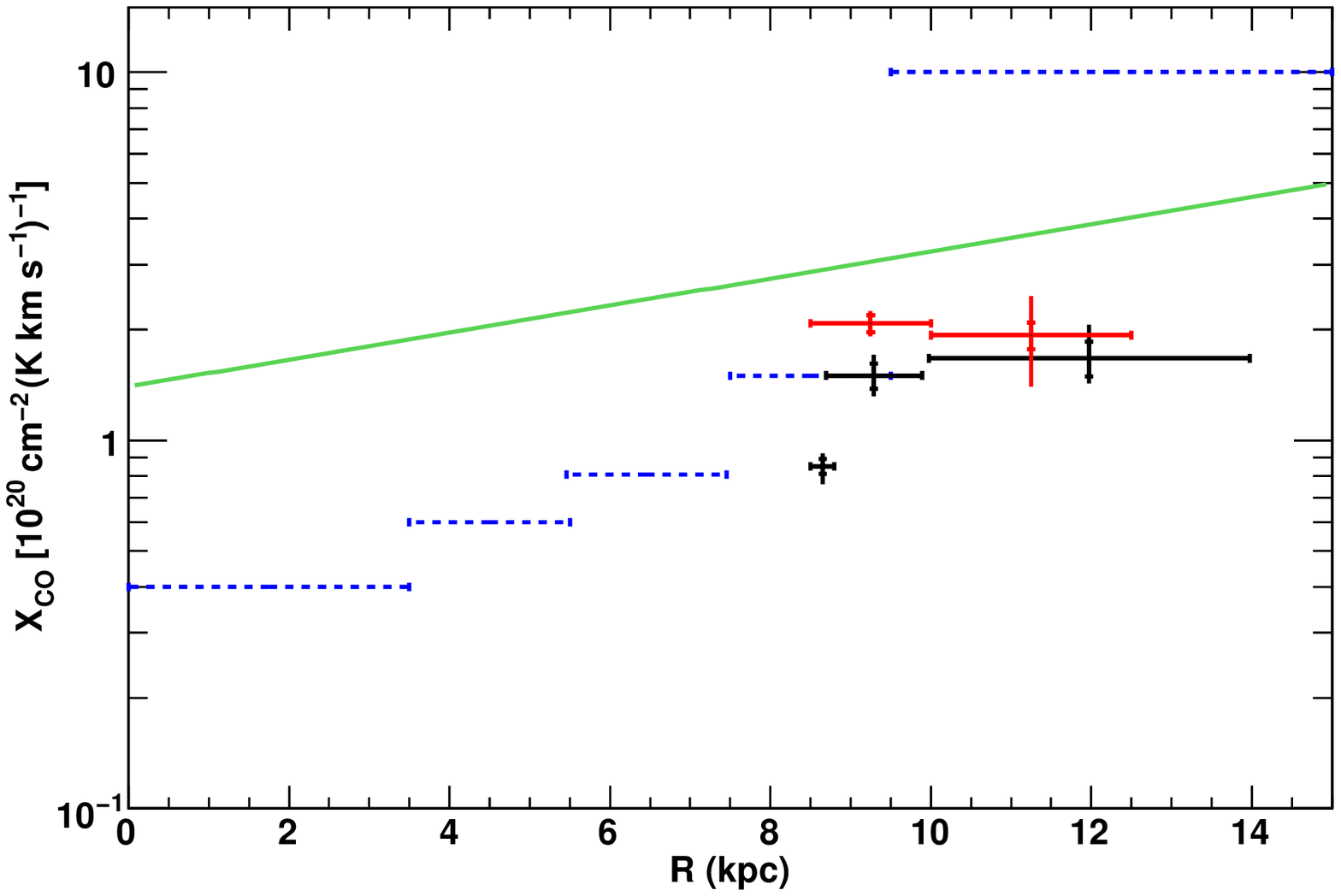}
\end{tabular}
\caption{Left: $\qhi$ (integrated above 200 MeV) as a function of Galactocentric
radius $R$ for different $\hi$ spin temperatures, as measured in the second (black) and third (red
or gray) quadrants. 
Statistical errors are comparable with marker sizes. 
Right: $\xco$ as a function of R. The error bars include
the uncertainties due to the $\hi$ optical depth correction. The (blue) step function
represents the model by Strong~\etal~\cite{ref:1}, the (green) line that by Nakanishi and
Sofue~\cite{ref:6}.}\label{fig:profiles}
\end{center}
\end{figure}

If subject to the same CR fluxes, a hydrogen molecule emits two times \g-rays as a
hydrogen atom. We
can therefore calibrate $\xco$ as $\qco/2\,\qhi$. Our results, constraining $\xco$
over a few~kpc from the solar system, are shown in fig.~\ref{fig:profiles} (right panel).

There is a significant increase by a factor~$\sim 2$ from the
nearby clouds of Cepheus and Cassiopeia in the Gould Belt to the
local arm, whereas no further significant variations are observed up to $R\sim14$~kpc. The
increase by one order of magnitude in the outer Galaxy proposed by Strong~\etal~\cite{ref:1} is not
confirmed. The \g-ray estimates are also systematically lower than predictions by Nakanishi and
Sofue~\cite{ref:6} from virial masses.

\section{The cosmic-ray gradient problem}

\begin{figure}[!bt]
\begin{center}
\begin{tabular}{cc}
 \includegraphics[width=0.5\textwidth]{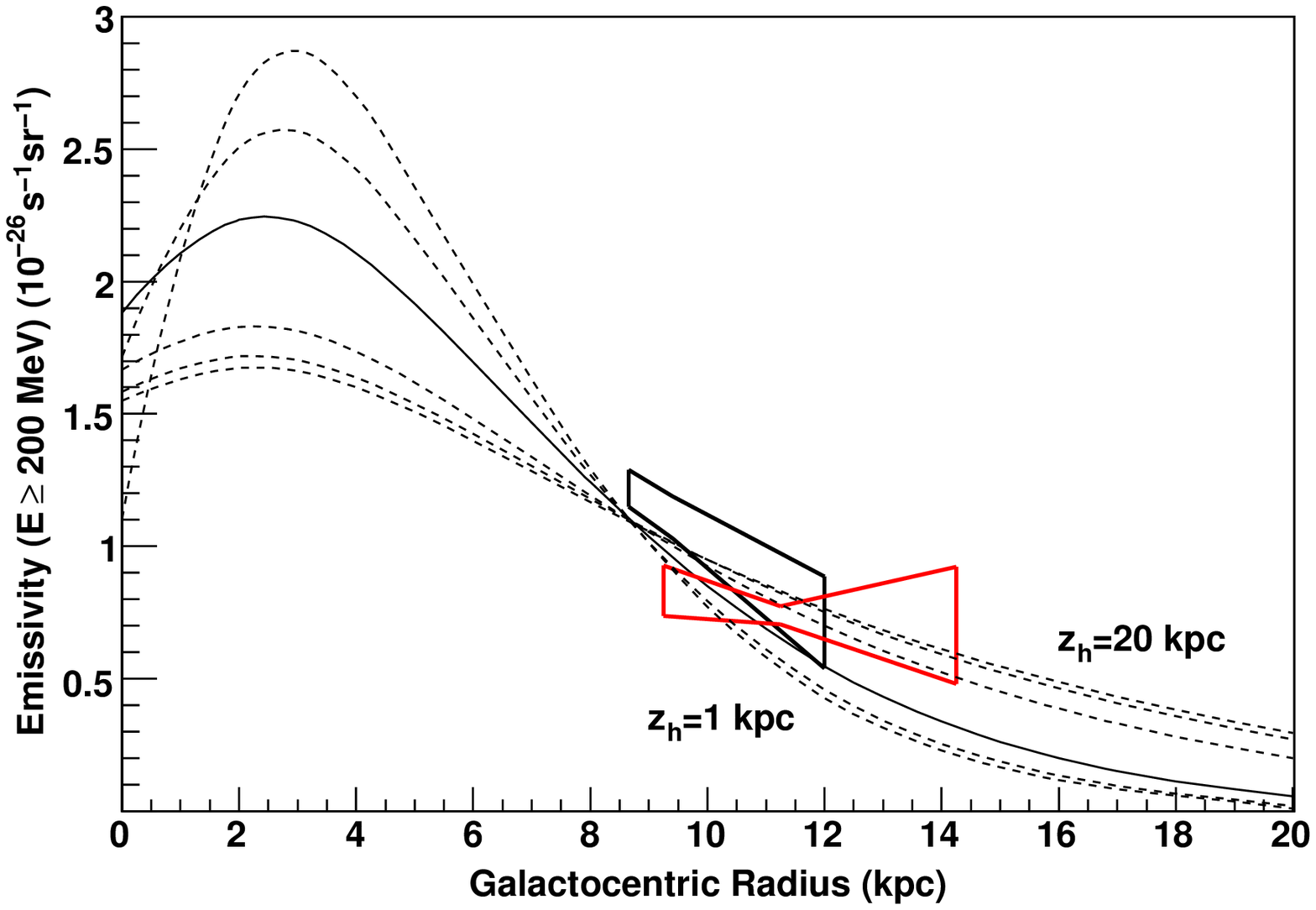} &
 \includegraphics[width=0.5\textwidth]{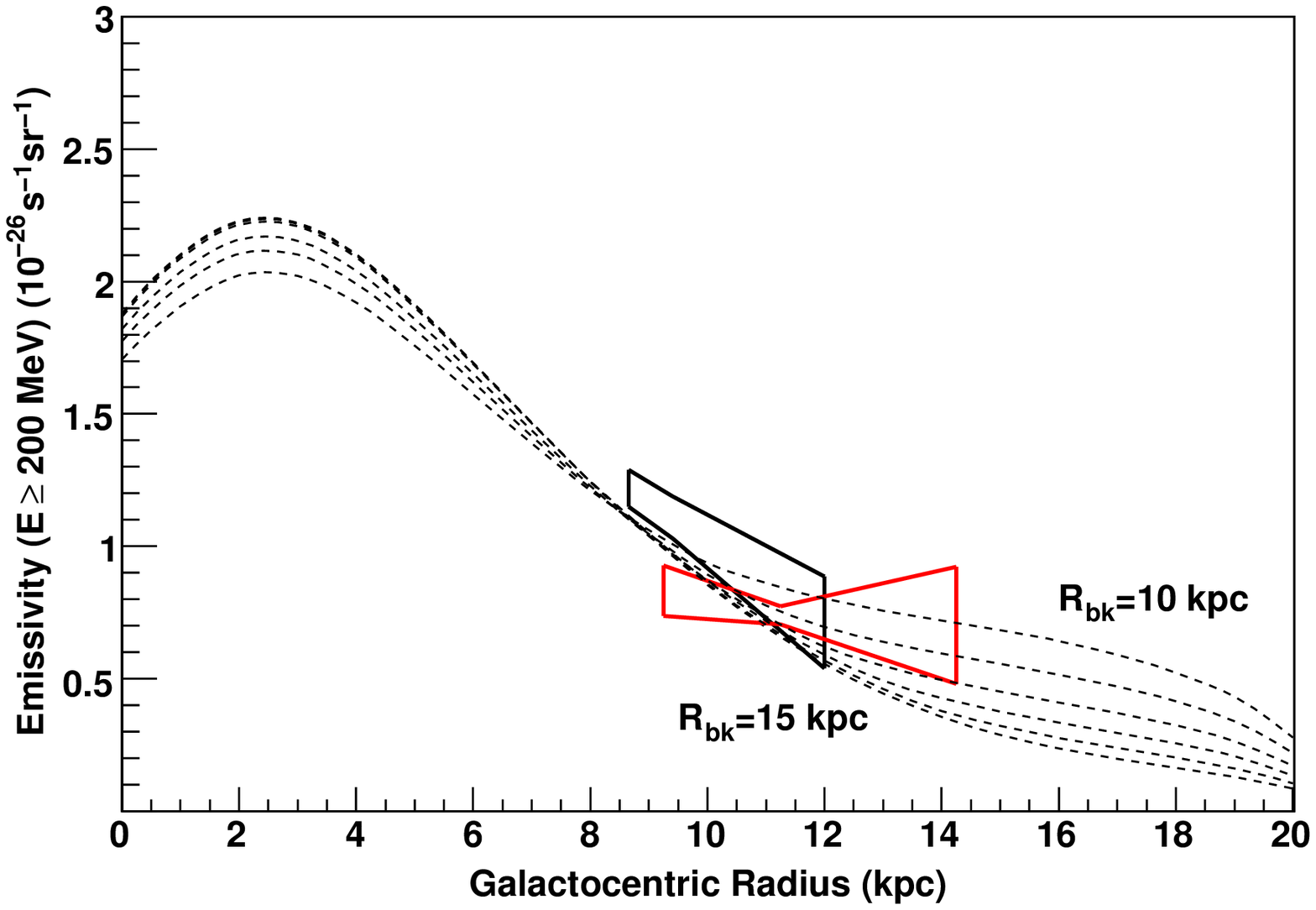}
\end{tabular}
\caption{$\qhi$ as a function of galactic radius $R$: bow-ties represent our estimates for the
second (black) and third (red or gray) quadrants. The curves give the predictions
by some GALPROP models, on the left varying the height of the propagation halo
from 1~kpc to 20~kpc (the solid line corresponds to 4~kpc, a value often assumed in past studies
\cite{ref:9}), on the right setting the density of CR sources to a
constant for $R>R_\mathrm{bk}=10-15$~kpc.}\label{fig:CRprop}
\end{center}
\end{figure}

The comparison between the measured \g-ray emissivities and expectations from CR propagation models has
implications for the origin and propagation of cosmic rays in the Galaxy. We adopted the widely-used GALPROP
propagation code
\cite{ref:8}. A conventional model analogous to
\cite{ref:1,ref:9} (solid line in fig.~\ref{fig:CRprop}, left panel) predicts a gradient steeper than inferred from the LAT measurements.

To alleviate the discrepancy one can increase the CR propagation halo or the
radial scale of the CR source distribution. 
In fig.~\ref{fig:CRprop} (left panel) we varied the halo height\footnote{And
correspondingly the diffusion parameters to stay consistent with CR isotopic abundances measured at
the
Earth.}, showing that halo heights from 10~kpc to 20~kpc are preferred if the CR source profile is
that derived 
from pulsar and supernova remnant observations. In fig.~\ref{fig:CRprop}
(right
panel) we assume a halo height of 4~kpc and we set the CR source density profile to a uniform
value beyond a given break radius. 
We found that \g-ray data point to a flat source distribution beyond $R \sim 10$~kpc.

The solutions are not unique: alternative CR propagation models can be considered, with,
\textit{e.g.}, a non uniform diffusion coefficient \cite{ref:10}. On the other hand, $\qhi$ might be
overestimated due to large amounts of dark gas in the outer
disc of the Milky Way not accounted for by radio/mm-wave data
\cite{ref:11}.

\acknowledgments
The \textit{Fermi} LAT Collaboration acknowledges support from a number of agencies and institutes
for both development and the operation of the LAT as well as scientific data analysis. These include
NASA and DOE in the United States, CEA/Irfu and IN2P3/CNRS in France, ASI and INFN in Italy, MEXT,
KEK, and JAXA in Japan, and the K.~A.~Wallenberg Foundation, the Swedish Research Council and the
National Space Board in Sweden. Additional support from INAF in Italy and CNES in France for science
analysis during the operations phase is also gratefully acknowledged.

\end{document}